\documentclass[a4paper]{article}

\usepackage{icrc2013}
\usepackage[english]{babel}
\usepackage{amsmath}

\newcommand{\en}{\varepsilon}
\usepackage{xcolor}

\title{Sensitivity of JEM-EUSO to Ensemble Fluctuations in the Ultra-High
  Energy Cosmic Ray Flux }

\shorttitle{Sensitivity of JEM-EUSO to Ensemble Fluctuations in the Ultra-High
  Energy Cosmic Ray Flux}

\authors{Markus Ahlers$^{1,2}$,
Luis A. Anchordoqui$^3$,
Angela V. Olinto$^{4,5}$,
Thomas C. Paul$^{3,6}$,
Andrew  M. Taylor$^7$
}

\afiliations{
$^{1}$ Wisconsin IceCube Particle Astrophysics Center (WIPAC), Madison, WI 53706, USA\\
$^2$ Department of Physics, University of Wisconsin, Madison, WI 53706, USA\\
$^{3}$ Department of Physics, University of Wisconsin-Milwaukee, Milwaukee, WI 53201, USA\\
$^{4}$ Department of Astronomy and Astrophysics, Enrico Fermi
Institute, University of Chicago, Chicago, Il 60637, USA\\
$^{5}$ Kavli Institute for Cosmological Physics,  University of Chicago, Chicago, Il 60637, USA\\
$^{6}$ Department of Physics, Northeastern University, Boston, MA 02115 USA\\
$^7$ Dublin Institute for Advanced Studies, 31 Fitzwilliam Place, Dublin 2, Ireland
}

\email{luis.anchordoqui@gmail.com}

\abstract{The flux of ultra-high energy (UHE) cosmic rays (CRs)
  depends on the cosmic distribution of their sources. Data from CR
  observatorions are yet inconclusive about their exact location or
  distribution, but provide a measure for the average local density of
  these emitters. Due to the discreteness of the emitters the flux is
  expected to show {\it ensemble fluctuations} on top of the statistical
  variations, a reflection of the {\it cosmic variance}.  This effect is strongest
  for the most energetic cosmic rays due to the limited propagation
  distance in the cosmic radiation background and is hence a local
  phenomenon. In this work we study the sensitivity of
  the JEM-EUSO space mission to ensemble fluctuations on the
  assumption of uniform distribution of sources, with local source
  density $\sim 10^{-5}~{\rm Mpc}^{-3}$. We show that in 3 years
  of observation JEM-EUSO will be able to probe ensemble fluctuations
  if the nearest sources are at 3 Mpc, and that after 10 years orbiting
  the Earth, this pathfinder mission will become sensitive to ensemble
  fluctuations if the nearest sources are 10 Mpc away. The study of
  spectral fluctuations from the local source distributions are
  complementary to but independent of cosmic ray anisotropy
  studies.
}

\begin{document}
\maketitle

\section{Introduction}

Above about 10~GeV, the CR spectrum falls as an approximate power-law
in energy, $dN/dE \propto E^{-\gamma}$. On a gross scale,
this power law appears rather featureless, but closer examination
reveals several breaks in the spectral index, $\gamma$.  The small
change from $\gamma \simeq 2.8$ to $\gamma \simeq 3.0$ at $E \approx
10^{6.5}~{\rm GeV}$ is known as the {\it knee}. The spectrum steepens
further to $\gamma \simeq 3.2$ above the {\it dip} ($E \approx
10^{8.7}~{\rm GeV}$), and then flattens to $\gamma \simeq 2.7$ at the
{\it ankle} ($E \approx 10^{10}~{\rm GeV}$).  The most recently
uncovered feature is a sharp and statistically very significant
suppression of the flux reported first in 2007 by the HiRes collaboration~\cite{Abbasi:2007sv},
and later confirmed by the Auger collaboration~\cite{Abraham:2008ru},
which reported $\gamma = 2.69 \pm 0.2 ({\rm stat}) \pm 0.06 ({\rm syst})$
and $\gamma = 4.2 \pm 0.4 ({\rm stat}) \pm 0.06 ({\rm syst})$ below and above
and $E = 10^{10.6}~{\rm GeV}$~\cite{Abraham:2008ru} respectively.
In 2010, an updated Auger measurement of
the energy spectrum was published~\cite{Abraham:2010mj}. The break
corresponding to the suppression is located at $\log_{10}(E/{\rm GeV})
= 10.46 \pm 0.03$.  Compared to a power law extrapolation, the
significance of the suppression is greater than $20\sigma$.\footnote{The existence of this suppression is
  consistent with the predictions of Greisen, Zatsepin and Kuzmin
  (GZK)~\cite{Greisen:1966jv,Zatsepin:1966jv}, in which CR
  interactions with the cosmic microwave background (CMB) photons
  rapidly degrade the CR energy, limiting the distance from which UHE
  CRs can travel to $\sim 100~{\rm Mpc}$.}

This collection of spectral features clearly reflects physically
interesting phenomena, including CR source distributions, emission
properties, composition and propagation effects. Indeed, there is a
large body of literature devoted to inferring such fundamental
information from details of spectral features. A common approach
involves developing some hypothesis about source properties and, using
either analytic or Monte Carlo methods, deducing the mean spectrum one
expects to observe at Earth. As our knowledge of source distributions
and properties is limited, it is common practice to assume spatially
homogeneous and isotropic CR emissions, and compute a mean spectrum
based on this assumption. In reality, of course, this assumption
cannot be correct, especially at the highest energies where the GZK
effect severely limits the number of sources visible to us. We can,
however, quantify the possible deviation from the mean prediction
based on the knowledge we do have on the source density and the
possible distance to the closest source populations.  This next
statistical moment beyond the mean prediction is referred to as
the {\it ensemble fluctuation}~\cite{Ahlers:2012az}. It depends on,
and thus provides information on, the distribution of {\em discrete}
local sources, source  composition, and energy losses during
propagation. This ensemble fluctuation in the energy spectrum is one
manifestation of the {\it cosmic variance}, which should also appear
directly through eventual identification of nearby source
populations. In fact, once statistics become sufficiently large, it will
be interesting to try to identify the ensemble fluctuations in the
energy spectrum in two 'realizations' of the universe; one could for
instance try to detect spectral variations in
the northern and southern skies which are consistent with ensemble
fluctuations, or, if a dipole anisotropy eventually
becomes evident, search for distinct ensemble fluctuations associated
with this spatial anisotropy. (JEM-EUSO searches for CR
anisotropy are discussed elsewhere in these proceedings~\cite{Tom}.)

It was recently pointed out that next generation experiments may be
sensitive enough to find evidence of this hitherto undetected ensemble
fluctuation at the high energy end of the
spectrum~\cite{Ahlers:2012az}.  Since the ensemble fluctuations
persist in the limit of large statistics, sufficiently large exposure
renders it feasible to discern them from statistical fluctuations in
the spectrum.  The magnitude and fine structure of ensemble
fluctuations depend on the density of UHECR sources, the composition
of the UHECR, and propagation effects.  For heavy nuclei, the
evolution of the spectra proceeds very rapidly on cosmic time scales
and the observed flux of secondary nuclei at Earth, $J(E)$, looks generally
quite different from the initial source injection
spectrum~\cite{Anchordoqui:1997rn,Ahlers:2011sd}.  As we will show
later, this fact has the potential to introduce distinctive shapes
into the energy spectrum.

Characterizing ensemble fluctuations is an exceedingly difficult task
owing to the rarity of the highest energy events. Space-based
observatories with huge exposures will be critical to pursue
this endeavor. The Extreme Universe Space Observatory on the Japanese
Experiment Module (JEM-EUSO) will be placed into orbit aboard the
International Space Station (ISS)~\cite{Adams:2012hr}.  For $E \geq 9
\times 10^{10}~{\rm GeV}$, this instrument will observe some $6 \times
10^4~{\rm km}^2 \, {\rm sr} \, {\rm yr}$
annually~\cite{Adams:2013vea}, about a factor of 9 greater than the
Pierre Auger Observatory nominal annual exposure (for zenith angles up
to $60^\circ$)~\cite{Blumer:2010zza}. In
this work we quantify the sensitivity of JEM-EUSO to ensemble
fluctuations.

\section{Calculation of ensemble fluctuations}

The calculation of ensemble fluctuations reflecting spatial variations that depend on the local
source density is formally divergent, unless we
introduce a regulator: $r_{\rm min}$. This mathematical regularization
has a physical interpretation as the distance to the closest source (or source population),
and it is these potential close by sources that introduce the
largest contribution to variations from the mean spectrum prediction.

\begin{figure}[!t]
  \centering
  \includegraphics[width=0.45\textwidth]{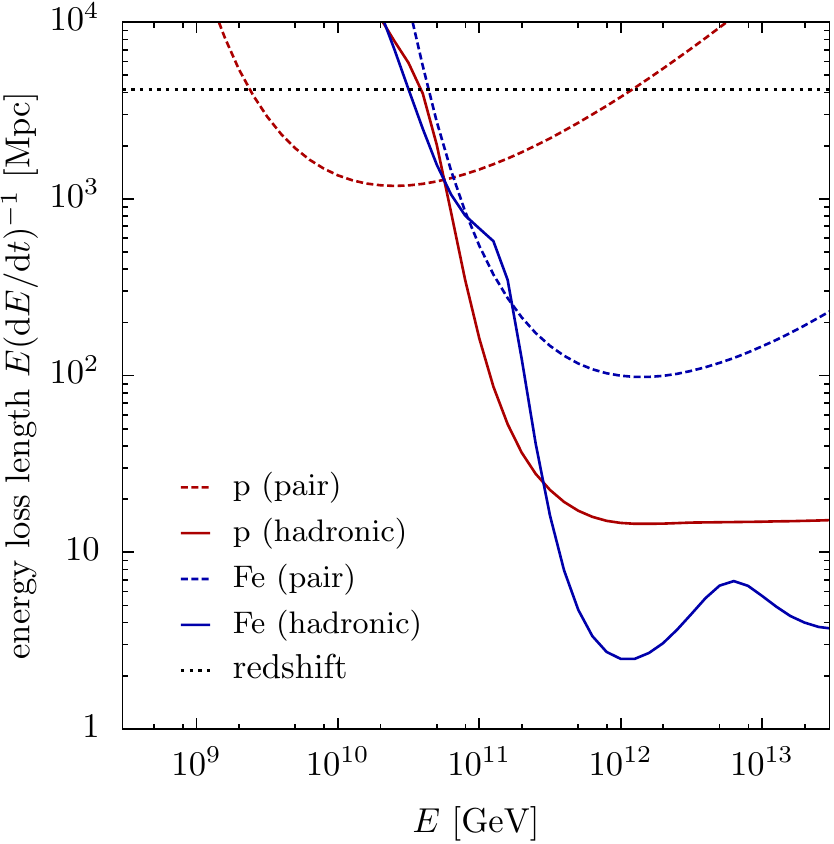}
  \caption{Energy loss length of $e^+e^-$ pair production (for proton
    and $^{56}$Fe nuclei), photopion production, $^{56}$Fe
    photodisintegration, and cosmological redshift.}
  \label{fig:uno}
 \end{figure}

As described in the Introduction, ensemble fluctuations are influenced
by various UHECR source and propagation characteristics, including
the distribution of {\em discrete} local sources, the nuclear
composition, and energy losses during propagation. If the primary CRs
are protons, the dramatic energy degradation proceeds via resonant
photopion production in the CMB. If the primary CRs are heavy
nuclei, successive photoevaporation of one or two nucleons through the
giant dipole resonance is mainly responsible for the UHECR energy
loss~\cite{Greisen:1966jv,Zatsepin:1966jv}. The energy loss length for
both of these processes is shown in Fig.~\ref{fig:uno}.  These severe
energy losses at high energies not only carve the famous GZK feature
at the end of the CR spectrum, but also strongly modulate
the structure of the ensemble fluctuation.

The secondary nuclei produced via photodisintegration carry
approximately the same Lorentz factor as the initial nucleus. It is
hence convenient to express the energy of a nucleus with mass number
$A$ as $A \en$ where $\en$ denotes the energy {\it per nucleon}.  The
differential interaction rate corresponding to the production of a
nucleus with mass number $B$ and energy $E'$ from a nucleus with mass
number $A>B$ and energy $E>E'$ can be approximated as $\gamma_{A\to
  B}(E,E') \simeq \Gamma_{A\to B}(E)\delta(E' - (B/A)E)$ where
$\Gamma_{A\to B}$ is the partial width of the transition.

Since, in this analysis, we are not concerned with the direction of the source we can
imagine sitting at the center of a sphere with radius $r_\star$, with
emission rate spectrum ${\cal Q}_A(E)/(4 \pi r_\star^2)$. The
point-source flux is described by Boltzmann's equation~\cite{Ahlers:2012az}
\begin{eqnarray}
\frac{1}{r^2}\partial_r (r^2 F_{A,i}) \!\!\!\!\! & \simeq &  \!\!\!\!\! \delta(r-r_\star)
\frac{Q_{A,i}}{4\pi
  r^2} \\
&  + & \!\!\!\!\! \Gamma^{\rm
  CEL}_{A,i+1}F_{A,i+1}-\Gamma^{\rm CEL}_{A,i}F_{A,i}  \nonumber  \\
&- & \!\!\!\!\!\!\sum_{B<A}\Gamma_{(A,i)\to(B,i)} F_{A,i}
 + \sum_{B>A}\Gamma_{(B,i)\to(A,i)} F_{B,i}, \nonumber
\label{eqn:boltz}
\end{eqnarray}
where $F_{A,i} \equiv \Delta \epsilon_i \ A\,{ dF}_{A}(A
\epsilon_i)/{d E}$ is the binned energy flux, $Q_{A,i} \equiv A \ \Delta
\epsilon_i \ \mathcal{Q}_A(A\epsilon_i)$ are the corresponding
emission rates, and $\Gamma_{(A,i)\to (B,i)} \equiv \Gamma_{A\to
  B}(A\epsilon_i)$.  Following~\cite{Ahlers:2010ty}, we adopt the continuous energy loss
(CEL) approximation to describe variations in the energy per nucleon,
$\Delta \epsilon$. The interaction
rates of CEL processes are  $\Gamma^{\rm CEL}_{A,i}  \equiv b_A(A
  \epsilon_i) /(A\Delta \epsilon_i)$, with $b \equiv d \epsilon/dt$.

\begin{figure*}[!t]
  \centering
  \includegraphics[width=\textwidth]{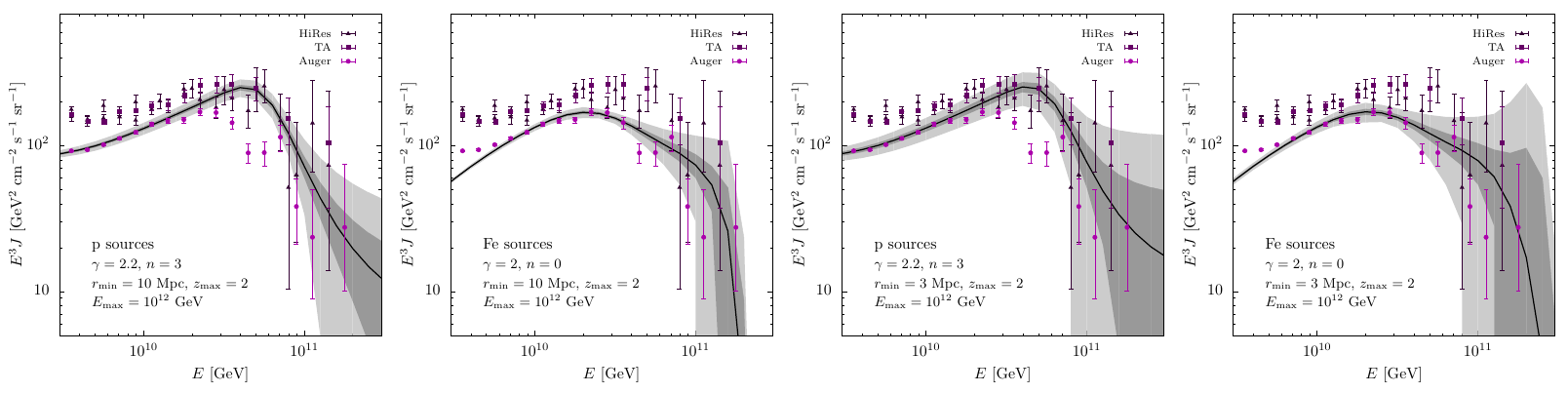}
  \caption{Approximate variation of the flux assuming a local source
    distribution ${\cal H}_0 = 10^{-5}~{\rm Mpc}^{-3}$ (dark gray
    band) ${\cal H}_0 = 10^{-6}~{\rm Mpc}^{-3}$ (light gray band) for
    different assumptions about primary species, emission spectral index
    $\gamma$, source evolution with redshift $(1+z)^n$, and $r_{\min}$. For the
    energy cuttof at the source, we have taken $E_{\rm max} =
    10^{12}~{\rm GeV}$.}
  \label{fig:dos}
 \end{figure*}

We want to study the statistical mean and variation of the aggregated flux of
$n_s$ local sources $N_{A,i} \equiv \sum_{s=1}^{n_s} F_{A,i} (r_s)$. Consider
$n_s$ sources distributed between redshift $r_{\rm min}$ and $r_{\rm max}$. The
number of sources can be expressed via a local source density $\mathcal{H}_0$ as
$n_s = \mathcal{H}_0 (4\pi/3)(r_{\rm max}^3-r_{\rm min}^3)$.  The probability
distribution function for a single source is $p(r) =
\frac{\mathcal{H}_0}{n_s}4\pi r^2\Theta(r-r_{\rm min})\Theta(r_{\rm max}-r)$.

The ensemble-average of a quantity $X(r_1,\ldots,r_{n_s})$, which depends on the distances of $n_s$ sources
can be expressed as $\langle X\rangle = \int {\rm d}r_1\cdot\ldots\cdot{\rm
  d}r_{n_s}p(r_1)\cdot\ldots\cdot p(r_{n_s}) X .$
Therefore, the ensemble-average of the local flux of particle species
$A$ is given by
$\langle N_{A,i}\rangle \equiv \mathcal{H}_0\int_{r_{\rm min}}^{r_{\rm
    max}} {\rm d}r'4\pi r'^2 F_{A,i}(r') $. The mean total flux is
obtained by summing over all particle species $\langle N_{\rm tot} (E) \rangle
  \equiv \sum_A \langle N_A(E/A) \rangle$. Now, defining the residual $\delta X \equiv X - \left<X\right>$, we can write the
covariance between the relative flux of two particle species $A, B$
populating energy bins $i, j$ :
$
\langle\delta N_{A,i} \delta N_{B,j}\rangle\equiv\langle N_{A,i}
N_{B,j}\rangle-\langle N_{A,i}\rangle\langle
N_{B,j}\rangle
$. The relative variation of the total flux is described by two-point density perturbations
\begin{equation}
\sigma^2_{\rm loc} = \sum_{A,B}\frac{\langle\delta N_{A}(E/A) \delta N_{B}(E/B)\rangle}{\langle N_{\rm tot}(E)\rangle^2}.
\end{equation}
$\sigma^2_{\rm loc}$ is the ensemble fluctuation, the sensitivity to which we are interested in exploring for
the case of JEM-EUSO.

\begin{figure}[!t]
  \centering
  \includegraphics[width=0.45\textwidth]{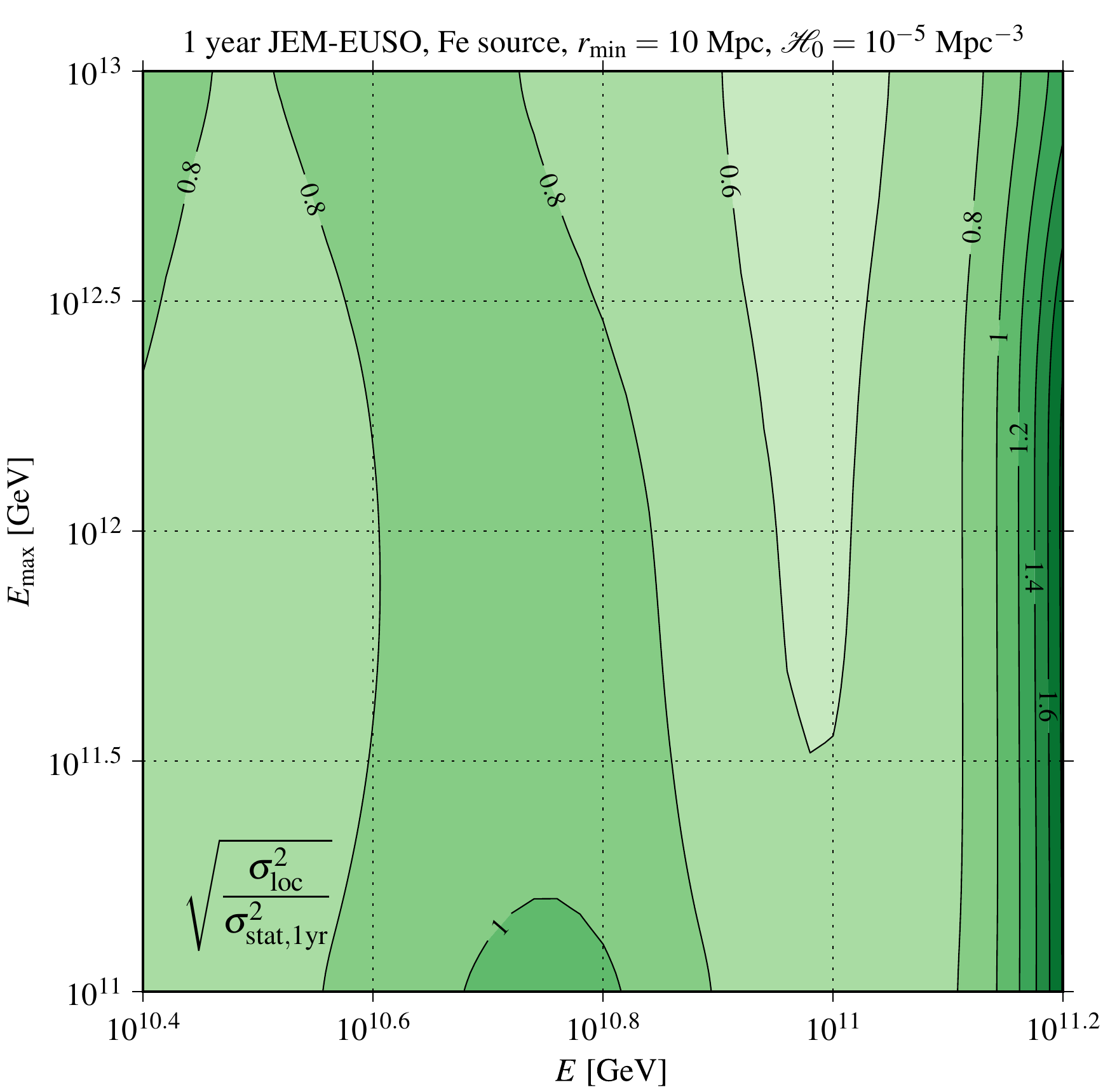}
  \caption{Contours of the ratio between the local relative error of the flux for a distribution of iron sources and the statistics to be collected by JEM-EUSO in one year. For the contour at 1, the statistical error equals the ensemble fluctuation ({\it i.e.} the cosmic variance is evident in the energy spectrum).}
  \label{fig:tres}
 \end{figure}

\section{Comparisons with recent data}

It is illustrative to compare the size and structure of ensemble
fluctuation to recent measurements. In Fig.~\ref{fig:dos}, we show
recent data from HiRes~\cite{Abbasi:2007sv},
Auger~\cite{Abraham:2010mj} and the Telescope Array
(TA)~\cite{AbuZayyad:2012ru} compared to the predictions derived from
Eq.~(\ref{eqn:boltz}) combined with a modification for a redshift
scaling out to $z=2$, as discussed in~\cite{Ahlers:2012az}. The solid
line shows the average energy spectrum predicted for certain assumed
parameters, which are indicated in the figures.  The dark gray band
envelopes the ensemble fluctuation for a uniform source density of
${\cal H}_0 = 10^{-5}~{\rm Mpc}^{-3}$, while the lighter band
corresponds to ${\cal H}_0 = 10^{-6}~{\rm
  Mpc}^{-3}$~\cite{Takami:2012uw,Abreu:2013wta}. The left two figures correspond to
the predictions for proton and iron, respectively, under the
assumption that $r_{\rm min} = 10$~Mpc~\cite{Taylor:2011ta}.  The
right two figures illustrate the situation for the case of $r_{\rm
  min} = 3$~Mpc. Note the striking dependence of the size of the
ensemble fluctuation on the proximity of the nearest source. In fact,
the rightmost figure indicates that for the case of iron primaries,
the ensemble fluctuation could even allow for a distinctive upward
wiggle in the spectrum before the GZK cutoff causes it to
plummet. (Candidate nearby sources of heavy nuclei include starburst
galaxies~\cite{Anchordoqui:1999cu} and ultra-fast spinning newly-born
pulsars~\cite{Blasi:2000xm,Fang:2012rx}.)

\begin{figure*}[!t]
  \centering
  \includegraphics[width=\textwidth]{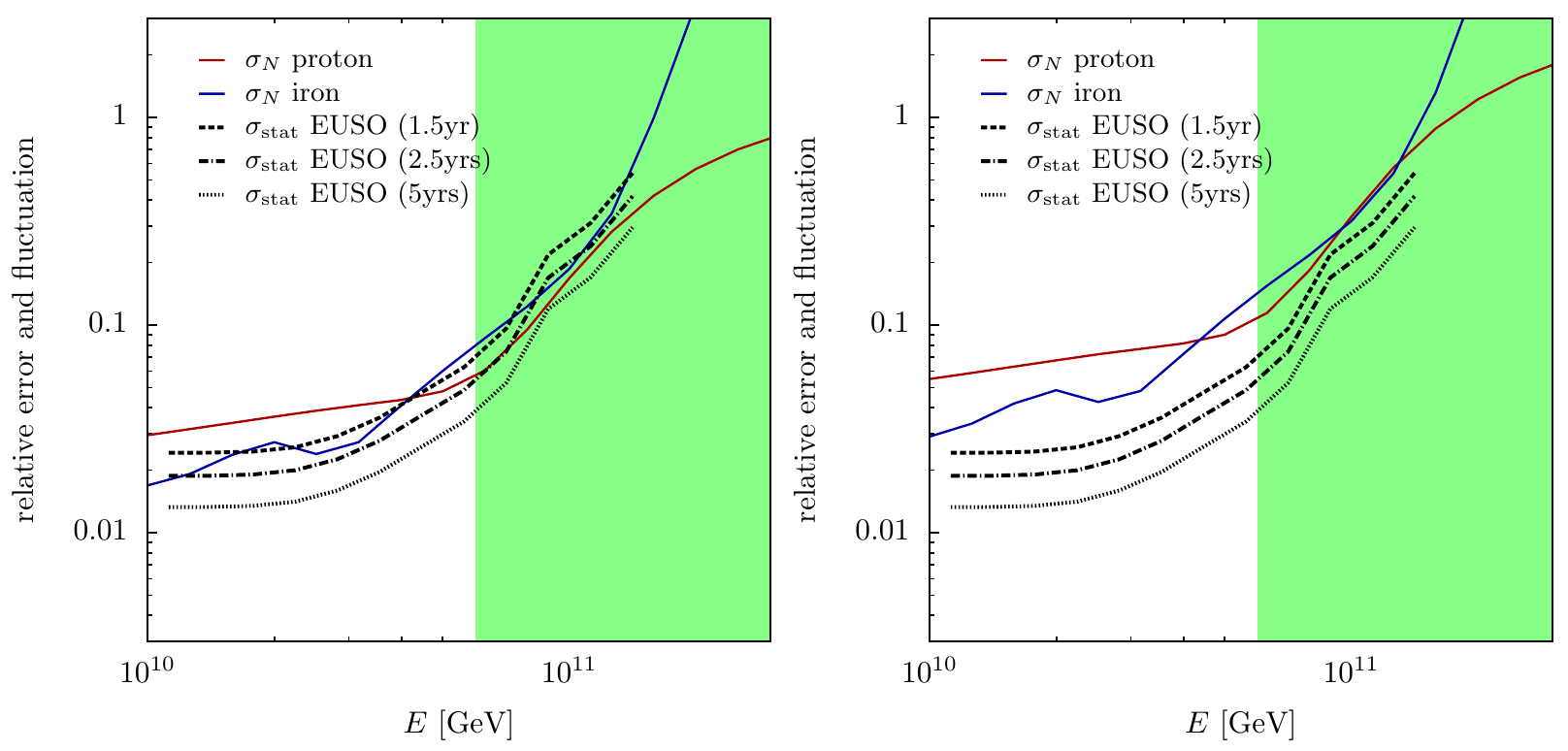}
  \caption{The relative error on the flux is compared to the
    statistical fluctuation for various integrated exposures for a
    slice through the contours in Fig~\ref{fig:tres} at $E_{\rm max} =
    10^{12}$~GeV. The left panel corresponds to the case of $r_{\rm
      min} = 10$~Mpc while the right panel corresponds to $r_{\rm min}
    = 3$~Mpc. The green shaded region is intended only to draw the
    reader's eye to the high energy region of the spectrum, where
    ensemble fluctuations are the largest.}
  \label{fig:cuatro}
 \end{figure*}

\section{JEM-EUSO discovery reach}

{}From Fig.~\ref{fig:dos}, it is evident that current volume of data at
the highest energies is not sufficient to disentangle the ensemble
fluctuation phenomenon from statistical fluctuations. Now we consider
whether the JEM-EUSO mission will attain sufficient exposure during
its projected lifetime to identify ensemble fluctuations for certain
assumptions about the source distribution and primary composition.

The Pierre Auger Observatory is currently the world's largest CR
detector.  Between 1 January 2004 and 31 December 2010 the observatory
collected an integrated exposure of ${\cal E} = 20,905~{\rm km}^2 \,
{\rm sr} \, {\rm yr}$~\cite{Abraham:2010mj}. For the JEM-EUSO baseline
design, a factor of 9 increase in annual exposure above $9 \times
10^{10}~{\rm GeV}$ is expected. To project the resulting JEM-EUSO
event rate, we scale the latest published Auger energy
spectrum~\cite{Abraham:2010mj} according to the relative exposure
between the Auger Observatory and
JEM-EUSO~\cite{Adams:2013vea}.\footnote{It is worth
    recalling that the Auger Collaboration currently reports a 22\%
    systematic uncertainty on the energy scale. If we normalize to the
    HiRes energy spectrum, JEM-EUSO statistics will increase by a factor of
    two.}

From this estimate of JEM-EUSO statistics, we find the expected ratio
of the ensemble to statistical fluctuations, $\sqrt{\sigma^2_{\rm
    loc}/\sigma^2_{\rm stat}}$. Figure~\ref{fig:tres} contains a
contour of this ratio in the plane of the maximum energy attained at
the source, $E_{\rm max}$ vs. the observed energy $E$.  The
precipitous rise in this ratio in the region of extremely high energy
is clearly evident.

Finally, in Fig.~\ref{fig:cuatro} we display
$\sqrt{\sigma^2_{\rm loc}}$ for the case of $E_{\rm max} =
10^{12}$~GeV compared to statistical uncertainties for various
JEM-EUSO exposure times and two different assumptions for $r_{\rm
  min}$.  The solid blue line indicates the relative fluctuation for
iron, while the red line corresponds to the proton fluctuation. Note
that JEM-EUSO is expected to collect data for at least 3 years, but
may attain significantly greater exposure (${\cal E} \approx 5 \times 10^5~{\rm
  km}^2 \, {\rm sr} \, {\rm yr}$) if ISS operations are extended beyond 2020.
Figure~\ref{fig:cuatro} demonstrates that if the nearest sources are
at about 3~Mpc, JEM-EUSO will gather enough statistics in 3 years to make an
observation of the ensemble fluctuation. Performing such an
observation in two 'realizations' of the universe, such as the
northern and southern skies, or regions of the sky demarcated by
anisotropy, will provide a measure of the cosmic variance which is
definitive and complementary to direct searches for the CR sources.
(It would perhaps be even more interesting to see no variation
in the energy spectra of two sky samples.)

\section{Conclusions}

Ensemble fluctuations are variations in mean energy spectrum predictions
reflecting the finite number and distribution of local sources.  This phenomenon
is one manifestation of the cosmic variance, and provides information
complementary to direct searches for the CR emitters.  The size and fine
details of ensemble fluctuations are sensitive to and therefore yield
information about the density of sources, the proximity to the nearest source or
source populations, and the composition of the highest energy CRs.
When taken together with information on CR clustering on a small angular
scale, the size of the ensemble fluctuation may also provide a lower bound
on the extragalactic magnetic field.

Ensemble fluctuations persist in the limit of large statistics, and may therefore
become evident given a sufficiently large data sample.  While current data sets
are not large enough to discern these fluctuations, future observatories with huge
exposure will afford an opportunity to turn up this phenomenon.  Here we have
shown that JEM-EUSO will collect sufficient statistics in 3 years of operation
to discern ensemble fluctuations manifest in two different samples of the sky if
the nearest sources are roughly 3~Mpc away. For more distant sources, at 10~Mpc,
10 years of running will reveal the ensemble fluctuations.

 \vspace*{0.5cm} \footnotesize{{\bf Acknowledgment:}{ We thank Etienne
     Parizot for some valuable discussion. This work was
     supported by NASA (11-APRA11-0058 and 11-APRA11-0066) and NSF
     (CAREER PHY-1053663, PHY-1068696, PHY-1125897, and PHY-1205845).

\end{document}